\providecommand{\pgfsyspdfmark}[3]{}

\documentclass[a4paper, 12pt]{article}

\usepackage{float}
\usepackage{graphicx}
\usepackage{booktabs}
\usepackage[utf8]{inputenc}
\usepackage{comment}
\usepackage{csquotes}
\usepackage{amsmath}
\usepackage{xcolor}
\usepackage{sectsty}
\usepackage{egothic}
\usepackage{marginnote}
\usepackage{tikz}
\usepackage{blindtext}
\usepackage{multirow}
\usepackage{makecell}
\usepackage{xcolor}
\usepackage{setspace}
\usepackage{breqn}
\usepackage{amsmath,amssymb}
\usepackage{geometry}
\usepackage[nobottomtitles*]{titlesec}
\usepackage{enumitem}
\usepackage{algorithm}
\usepackage[noend]{algpseudocode}

\usepackage{colortbl}

\usepackage{array}
\newcolumntype{P}[1]{>{\centering\arraybackslash}m{#1}}
\newcolumntype{L}[1]{>{\raggedleft\arraybackslash}m{#1}}
\newcolumntype{R}[1]{>{\raggedright\arraybackslash}m{#1}}

\usepackage{caption,subcaption,graphicx}

\usepackage{footnote}
\makesavenoteenv{tabular}

\egothfamily

\usepackage[
backend=biber,
style=alphabetic,
maxnames=50
]{biblatex}
\definecolor{PicGreen}{RGB}{0,115,115}

\usepackage[english]{babel}

\usepackage{longtable}
\title{
Deep differentiable reinforcement learning and optimal trading}
\author{Thibault Jaisson\footnote{The author thanks Reda Messikh, St\'ephane Daul and R\'emy Cottet for useful discussions on the topic and feedbacks on the manuscript and two anonymous referees for their remarks and recommendations.} \footnote{The opinions expressed in this article are solely those of the author.}\\
Pictet Asset Management\\
tjaisson@pictet.com}

\begin{document}


{\let\newpage\relax\maketitle}

\begin{center}
\textbf{Abstract}
\end{center}

\noindent In many reinforcement learning applications, the underlying environment reward and transition functions are explicitly known differentiable functions. This enables us to use recent research which applies machine learning tools to stochastic control to find optimal action functions. In this paper, we define differentiable reinforcement learning as a particular case of this research.
We find that incorporating deep learning in this framework leads to more accurate and stable solutions than those obtained from more generic actor critic algorithms.
We apply this deep differentiable reinforcement learning (DDRL) algorithm to the problem of one asset optimal trading strategies in various environments where the market dynamics are known. Thanks to the stability of this method, we are able to efficiently find optimal strategies for complex multi-scale market models. We also extend these methods to simultaneously find optimal action functions for a wide range of environment parameters. This makes it applicable to real life financial signals and portfolio optimization where the expected return has multiple time scales.
In the case of a slow and a fast alpha signal, we find that the optimal trading strategy consists in using the fast signal to time the trades associated to the slow signal.

\vspace{10pt}

\noindent \textbf{Keywords:} Reinforcement learning, Optimal trading, Differentiable, Stochastic control, Dynamic programming, Deep learning, Multi-scale signals.

\section{Introduction}

\noindent Reinforcement learning is a field of machine learning which aims at finding optimal sequences of actions in order to achieve a multi-step task or maximize a cumulative reward.
Its combination with deep learning, see \cite{lecun2015deep}, called deep reinforcement learning (DRL), see \cite{franccois2018introduction}, has recently  had a huge success in addressing sequential decision problems.
For example, it outperforms the best humans at many games, see \cite{mnih2015human}, and has state of the art applications in many real life applications such as  robotics, see \cite{lillicrap2015continuous}, self-driving cars, see \cite{pan2017virtual} and finance, see \cite{cartea2021deep}.\\

\noindent Quantitative portfolio management is the use of mathematical and statistical tools to devise `optimal' portfolios of financial assets.
Historically, this field mostly focused on getting the best expected return (or alpha in the following) with the smallest possible risk as in the mean variance framework of Markowitz, see \cite{markowitz1968portfolio}.
However, this does not take into account the practically paramount role of trading costs: An investor needs to control his turnover if he does not want to spend his alpha in trading costs, see \cite{frazzini2012trading}.
This implies that the current choice of portfolio weights does not only impact the portfolio returns over the next short period but also the future investor positioning.
This naturally puts us in the setting of reinforcement learning where the current actions (trades) have an impact on the next reward as well as on the future state of the environment which itself has an impact on future rewards, see \cite{sutton2018reinforcement}.\\

\noindent Over the last decade, the use in finance of modern machine learning techniques such as deep learning has seen a huge increase to: price options, see \cite{horvath2019deep}, predict returns, see  \cite{gu2018empirical} and model risks, see \cite{gu2019autoencoder}.
In spite of the challenges inherent to financial data such as the structural low signal to noise ratio and non stationarity, adding non linearity and interactions between usual features seems to add value compared to more classical linear models, see \cite{daul2021performance}.
Deep reinforcement learning has also been extensively applied to directly build optimal strategies from real financial data, see \cite{deng2016deep} and \cite{cong2020alphaportfolio}, by minimizing some loss of the portfolio performance.
These approaches have the particularity that they do not split the portfolio construction process in two steps: First modeling and estimating the dynamics of the market (which is typically a data science issue with no mathematically perfect solution). Second finding the optimal strategy in this model (which is an engineering problem with a mathematically well defined solution).
This however makes it harder for an investor to understand his portfolio and what criteria explain the asset weights.\\

\noindent An alternative approach taken in \cite{chaouki2020deep}, which we follow here, is to assume that the market is well modeled and its parameters are known. In this model, we can then use deep reinforcement learning to find the optimal strategy. The advantage of their approach is that one can always simulate as much data as one needs and evaluate strategies as precisely as necessary. 
They use the state of the art Deep Deterministic Policy Gradient (DDPG). This algorithm tends to be hard to train and requires much engineering, meta-parameters tweaking and to be trained on different starting points to be well fitted in practice, see \cite{fujimoto2018addressing}. Even then, it is only applied to a very simple $AR(1)$ alpha model and when more complex models are given as input, the strategies obtained do not converge towards the optimal ones.\\

\noindent In particular, in this article, we are interested in devising optimal trading strategies in environments where the alpha process has a multi-time scale behavior, see \cite{garleanu2013dynamic} and \cite{boyd2017multi}. Indeed, in practice, investors have access to return predictors with different reversion time scales.
For example, Value, Momentum, see \cite{asness2013value}, and Quality, see \cite{asness2019quality}, can be considered as slow signals, indeed if a stock has a good Quality signal it will most likely remain good in the coming months. On the other hand, price reversal, see \cite{bremer1991reversal}, and analyst recommendations, see \cite{jaisson2021predictive}, are much faster signals with shorter memory. We tried to apply the DDPG algorithm to such multi-scale trading environments however, even after many attempts, we did not manage to make it converge to optimal strategies.\\

\noindent To solve this issue, we follow recent research which applies machine learning tools to stochastic control problems, see \cite{fecamp2019risk}, \cite{germain2021neural} and \cite{han2016deep}. It leverages on the fact that many  reinforcement learning problems can be cast as optimal control where the underlying dynamics of the environment are known. This enables to precisely anticipate the future impact of actions and to directly maximize the cumulative expected reward.
Looking for action functions which take the form of a feed forward neural network (FFNN) gives an algorithm which derives optimal trading strategies in the mono-asset case. We call  this framework deep differentiable reinforcement learning.
Furthermore we thoroughly analyze and describe the interactions between the slow and fast signals in the optimal action. We find that the slow signal is more correlated to the portfolio weights and the fast signal is more correlated to the trades. In other words, the fast signal is used to time the trades associated to the slow signal.\\


\noindent This paper is organized as follows:
In Section \ref{sec:drl},  we introduce the differentiable reinforcement learning framework and a generic algorithm to find optimal action functions approximated by a deep neural network.
In Section \ref{sec:opt_trade}, we show how generic (multi-scale) optimal trading problems fit into the context of differentiable reinforcement learning.
In Section \ref{sec:num_res}, we exhibit and describe the properties of optimal trading strategies. We conclude on how to apply this in practice in Section \ref{sec:conc}.

\section{Differentiable reinforcement learning}
\label{sec:drl}

\subsection{The reinforcement learning framework}

In the standard reinforcement learning setup, see \cite{sutton2018reinforcement}, a learning agent is interacting with his environment in discrete time in the following way: At each time $t$, he sees the current sate of the environment $s_t$ (in some settings the agent only sees a noised version of the environment but this is irrelevant in what follows) and chooses an action $a_t$. The environment then samples a reward $r_t$ to the agent and shows it the next state $s_{t+1}$.
The aim of the agent is to find a sequence of actions which maximize the expectation of the cumulative reward up to time $T$ which is defined as
\begin{equation}
\textrm{CR}_T= \sum_{i=0}^{T-1}r_{i}.
\end{equation}

\noindent What makes the problem challenging is that both the reward and the next state (and thus future rewards) depend on the actions of the agent which thus have short term and long term impact.\\

\noindent Note that in some reinforcement learning settings, the objective of the agent is to maximize the  infinite sum of discounted future rewards, see \cite{lillicrap2015continuous}. If the discounting factor is close to one and $T$ is large, the optimal action functions for these two objectives should be very similar. However, as we will see, having a large but finite sum enables us to minimize it efficiently.\\

\noindent A noteworthy point is that as opposed to the more classical optimal control setting, see \cite{bertsekas2012dynamic}, in the reinforcement learning setting, the underlying dynamics of the environment are not necessarily known by the agent. From his point of view, the environment is simply a black box from which we can sample state transitions and rewards by giving it actions.\\

\noindent Until recently, the choice of an optimal action function was only solvable either in very specific settings where a closed form solution exists or where both the action and the state spaces are finite and rather small, see \cite{watkins1992q}. In the last decade, the combination of ideas from optimal control with the flexibility of deep learning has enabled astonishing progress in the context of self-driving cars, see \cite{pan2017virtual}, games, see \cite{mnih2015human} and robotics, see \cite{lillicrap2015continuous}.
First for discrete action spaces, see \cite{mnih2015human}, and even more recently for continuous and potentially high dimensional action spaces. For example \cite{lillicrap2015continuous} introduced the Deep Deterministic Policy Gradient (DDPG) algorithm which jointly learns the optimal action and the value function by approximating them with deep neural networks and using the Bellman equation.\\

\noindent However, even if progress is made every year in this direction, see \cite{fujimoto2018addressing}, this kind of methods called  actor critic are still hard to train. This is especially the case when the expected cumulative reward is flat around the optimal action where they require much tweaking of the meta-parameters and engineering of the algorithm, see \cite{ioffe2015batch} and \cite{henderson2018deep}. Even then, the results are far from perfect and need to be trained on many learning paths with different starting points to get closer to the optimal solution, see \cite{chaouki2020deep} and Appendix \ref{app:ddpg}.\\

\noindent In the next paragraph, we introduce the differentiable reinforcement learning setting. It is very close to that of recent works which apply neural networks to solve stochastic control problems, see \cite{fecamp2019risk}, \cite{germain2021neural} and \cite{han2016deep}. These approaches use the fact that in many applications such as the one of \cite{chaouki2020deep}, the underlying dynamics of the environment are known. Moreover, in a sense that we will make precise, these dynamics are differentiable. In this setting, we follow these works to devise a conceptually simple algorithm which finds optimal action functions that are far more accurate and stable than the ones obtained from generic reinforcement learning methods, see Appendix \ref{app:ddpg}.\\

\noindent Applied to the problem of optimal trading strategies, it enables us to optimize more complex market models with a large number of meta-parameters.

\subsection{The differentiable reinforcement learning framework}

We define the differentiable reinforcement learning framework as a particular case of discrete time stochastic control:

\begin{itemize}
\item  The next state $s_{t+1}$ is a known differentiable transition function $\Phi$ of the current state $s_t$, the chosen action $a_t$ and a random variable $U_{t+1}$ that we can sample from
\begin{equation}
s_{t+1} = \Phi(s_t, a_t, U_{t+1}).
\end{equation}
\item  The reward $r_t$  is a known  differentiable function $\Psi$ of the current state $s_t$, the chosen action $a_t$ and a random variable $V_{t}$ that we can sample from
\begin{equation}
r_t = \Psi(s_t, a_t, V_t).
\end{equation}
\end{itemize}

\noindent In this framework, the variables $U$ and $V$ fully capture the randomness of the environment.\\

\noindent The aim of optimal control problems is to find an action function $A$ such that $$a_t=A(s_t)$$which maximizes the expectation of the long term cumulative reward. Note that in  \cite{fecamp2019risk} and \cite{han2016deep}, the authors either fit one function per time to maturity of take the time to maturity as an input of the action function. In the following, we consider `stationary' environments where the horizon $T$ is not a real objective but a technical parameter that just needs to be large enough compared to the environment time scales. To reduce the input dimension of the action function, we do not take time to maturity as input. \\

\noindent Note that applications that take as input real world data instead of simulated ones, do not make it into this framework. However, when working with known environments that we can simulate we are mostly in this setting.

\subsection{The cumulative reward function}

An important property of this framework is that given an action function $A$ the cumulative reward $\textrm{CR}_T$ of the strategy is only a function of $A$, $s_0$, $(V_t)_{0\leq t\leq T -1}$ and $(U_t)_{0\leq t< T-1}$:
\begin{equation}
\textrm{CR}_T = \textrm{CR}_T(A, s_0,(V_t)_{0\leq t\leq T -1},(U_t)_{0\leq t< T-1}).
\end{equation}
For example, the rewards for T=1, 2 and 3 are shown below.
\begin{footnotesize}
\begin{itemize}
\item $\textrm{CR}_1=\Psi(s_0, A(s_0), V_0).$
\item $\textrm{CR}_2 = \textrm{CR}_1+\Psi(\Phi(s_0, A(s_0), U_0), A(\Phi(s_0, A(s_0), U_0)), V_1).$
\item $\textrm{CR}_3 = \textrm{CR}_2+\Psi(\Phi(\Phi(s_0, A(s_0), U_0), A(\Phi(s_0, A(s_0), U_0)), U_1), A(\Phi(\Phi(s_0, A(s_0), U_0),\\  A(\Phi(s_0, A(s_0), U_0)), U_1)), V_2).$
\end{itemize}
\end{footnotesize}

\noindent Of course for $T$ large, this function is quite deep in the sense that it is a composition of hundreds of base functions. However, we will see that we are able to efficiently minimize its expectation with respect to the action function for large $T$ (up to at least 100 which is more than enough for our applications).

\subsection{Deep learning application}

In order to numerically find the action function which maximizes the expectation of the cumulative reward up to a given time $T$, we need to restrict it to a family of parametrized action functions, with parameters denoted $\theta$: 

\begin{equation}
A(s)=F(s;\theta).
\end{equation}
Assuming that $F$ is differentiable with respect to these parameters $\theta$, the cumulative reward samples are also differentiable  with respect to $\theta$.
To numerically compute the associated derivatives of these deep composition of functions, we use automatic differentiation (AD) and backpropagation through the Pytorch package, see \cite{paszke2017automatic}. 
A stochastic gradient descent (SGD) like algorithm can then be used to find the action parameters which minimize the empirical sample average of the cumulative reward.\\

\noindent A flexible extensively studied and numerically convenient parametrization of the action function used is a similar manner in \cite{fecamp2019risk} is the dense feed forward neural network or multi-layer perceptron, see \cite{schmidhuber2015deep} for the definition. In what follows, unless stated otherwise, we take a neural network with two hidden layers with 300 hidden neurons each and $ReLU$ activation functions so that the model has around 10'000 parameters.

\subsection{Sampling and empirical average}

We fit $A$ (or equivalently $\theta$) by minimizing a loss function defined as the opposite of the empirical average of the cumulative reward on $N$ independent samples of  initial states $s_0^i$ and randomness $(U^i_t)_{t\leq T}$ and $ (V^i_t)_{t\leq T}$:

\begin{equation}
\min_\theta -\sum_{i=1}^N \textrm{CR}_T(F(\cdot, \theta), s^i_0, (V^i_t)_{0\leq t\leq T}, (U^i_t)_{0\leq t\leq T-1}).
\end{equation}

\noindent Note that the sampling distribution of $s_0^i$ does not matter too much. What is important is that the sample covers the states that we are interested in.\\

\noindent In what follows, unless stated otherwise, we minimize the above loss function with $T=50$ taking $N=10$ million samples with the Adam algorithm, see \cite{kingma2014adam}, with 1024 mini-batch sizes and the default parameters of the Pytorch library: lr=0.001 and betas=(0.9, 0.999), see \cite{paszke2017automatic}. We perform 50 epochs on the entire sample size. At each epoch, we reduce the learning rate by 10\% to improve convergence. See Appendix \ref{app:code} for the pseudo code.\\

\noindent The complexity of the model is
$$\#\{\text{model parameters}\}\times T \times N \times \text{epochs}.$$
On a a server using 16 \textit{2.9 GHz Xeon Platinium 8268} CPUs it takes around 10 hours to train the model (this time is weakly dependent on the environment complexity). Note however that for the simplest environments, we get a very good solution after the first epoch.

\subsection{Making environment parameters variable}

To make it usable in practice in a quantitative strategy, we need to be able to apply this to potentially thousands of stocks with different environment parameters (costs, spread, max weights, ...). It would not be efficient to fit a model for each set of parameters. One easy way to solve this issue is to consider these environment parameters as static (that is non time varying) parts of the state.
We therefore only need to train one model which takes as input the state and the environment parameters to simultaneously find the optimal action for a continuum of models.\\

\noindent More formally denote $\zeta$ an environment parameter:
$$s_{t+1} = \Phi(s_t, a_t, U_{t+1};\zeta) \text{ and }r_t = \Psi(s_t, a_t, V_t;\zeta).$$
Then defining the new state as $s'_t=(s_t,\zeta)$, we are in the same context as before without environment parameters:
$$s'_{t+1} = \Phi'(s'_t, a_t, U_{t+1}) \text{ and }r_t = \Psi'(s'_t, a_t, V_t).$$
We can thus find the optimal action as a function of $s_t$ and $\zeta$.\\

\noindent As for initial states, the environment parameter sampling does not matter too much and must simply cover the range of environments that one is interested in.

\section{Conception of optimal trading strategies}
\label{sec:opt_trade}
\subsection{Optimal investment and reinforcement learning}

As discussed in the introduction, when choosing his weights the investor must not only consider his immediate reward but also how he will be positioned in the future. As stated in \cite{garleanu2013dynamic}: `A good hockey player plays where the puck is. A great hockey player plays where the puck is going to be'.\\

\noindent Three papers which effectively tackle this issue are: \cite{garleanu2013dynamic}, where the authors find a simple intuitive solution in the simple case where the transaction costs are quadratic and related to the covariance matrix in a strong way. \cite{boyd2017multi}, where the authors use convex optimization in a multi-period setting to find a heuristic solution by assuming that the future alpha trajectory is deterministic. As we will see, this assumption yields sub-optimal strategies when the alpha is stochastic. \cite{de2012optimal}, where the authors use optimal control theory to find the optimal weight trajectory when costs are linear.\\

\noindent Another branch of application of deep reinforcement learning in finance is illustrated in \cite{cong2020alphaportfolio}. Instead of splitting the model estimation and the conception of an optimal strategy in this model in two steps, it consists in giving as input to a large carefully engineered neural network the raw features and expect as output optimal weights. See \cite{kolm2020modern} for a recent review of the applications on reinforcement learning in quantitative finance.\\

\noindent In this article, we follow the more classical road which consists in working in two steps:
\begin{itemize}
\item{First modeling (and estimating) the expected return (for example using supervised machine learning as in \cite{gu2018empirical}), risk and costs related to our assets.}

\item{Then find the optimal strategy under in this model.}
\end{itemize}

\noindent Although the two approaches have merits, some advantages of splitting modelization and optimization is that it makes it easier to: decompose our performance, test our assumptions, have margin of errors on our alpha estimations and make active decisions. Moreover, in this setting, the strategy conception step is not anymore a data estimation problem (since we can always simulate more data from the model) but an optimization problem whose tentative solutions can be precisely evaluated.\\

\noindent We are thus in the footsteps of \cite{chaouki2020deep}. However we want to work with more complex alpha models. In such models, we did not manage to make our action functions converge with the DDPG algorithm. In particular, we want to consider models where our alpha process has (at least) two time scales: a slow and a fast.  To illustrate these alphas, assume that the slow alpha has a half-life of one year (think Value, Momentum, Quality, see \cite{asness2013value} and \cite{asness2019quality}) and the fast alpha has a half-life of one month (think short term price reversal, analyst revisions, sector momentum, see \cite{bremer1991reversal} and \cite{jaisson2021predictive}).\\

\noindent The kind of non-trivial properties that we hope to find is that for reasonable model parameters, the fast alpha times the execution of the slow alpha. Indeed, assume that we expect a stock to perform well in the next year but not in the next month. If we do not have the stock in our portfolio, we will wait a month before buying it. If we have it in our portfolio, it might not be worth to sell it and to buy it again in a month because of transaction costs.

\subsection{Reward and objective}

\label{subsec:reward}

As in \cite{chaouki2020deep}, we will consider environments where the  reward writes as a combination of a return term equal to the return times the position, a risk term and a cost term:
\begin{equation}
r_t=w_t R_{t+1} -  \text{Risk}(w_t) - \text{Cost}(w_t, lw_t)
\end{equation}
where $R_{t+1}$ is the asset return over the next period, $w$ is the weight to be chosen and $lw$ is the last weight before the choice of the new weight. Assuming that the return writes as:
$$R_{t+1}= \alpha_t+\sigma N^R_{t+1}$$
where the expected return $\alpha_t$ is observable at time $t$, $\sigma$ is the stock volatility and $N^R_{t+1}$ are independent standard normal variables, an equivalent problem is to maximize the expectation of the cumulated expected reward defined as:
\begin{equation}
r_t=w_t \alpha_t -  \text{Risk}(w_t) - \text{Cost}(w_t, lw_t).
\end{equation}
\noindent In Appendix \ref{app:noise}, we study the impact of considering the `noised' reward on the convergence speed but in what follows, we consider this expected reward.\\

\noindent In this case, the state is the alpha and the last weight:$$s_t=(\alpha_t, lw_t)$$and the action is the next weight $w_t$.\\

\noindent In what follows, the action has a trivial effect on the next state: The last dimension of the next state is equal to the action. However, we can imagine trading environments where the action has a more complex effect on the alphas at different scales which is equivalent to the problem introduced in \cite{curato2017optimal} where market impact is transient and the kernel is approximated by a combination of exponential functions.

\subsection{Trading environments description}

In this paragraph, we describe the different trading environments that we will test. We give examples of alpha processes, risk functions and cost functions.\\

\noindent Note that all the combinations of models defined below enter the differentiable reinforcement learning framework.

\subsubsection{Alpha modeling}

Let us begin by defining the two alpha dynamics that we will consider: mono-scale and multi-scale.

\paragraph{Mono-scale alpha}

In this model, as in \cite{chaouki2020deep}, the alpha of the asset is an observable $AR(1)$ process (with $\rho_\alpha =0.9$ and $\eta_\alpha =1$ for example):
\begin{equation}
\alpha_t=\rho_\alpha \alpha_{t-1} + \eta_\alpha N^\alpha_t
\end{equation}
where $N^\alpha_t$ are independent standard normal variables.\\

\noindent The parameter $\eta_\alpha$ should be seen as a alpha scaling parameter. Indeed, the distribution of the $\alpha_t$ is a Gaussian of mean zero and standard deviation $\eta_\alpha / \sqrt{1-\rho_\alpha^2}$.

\paragraph{Two-scale alpha}

In this model, the alpha of the asset is a combination of two (independent) observable $AR(1)$ processes, a fast (F) and a slow (S) (with $\rho_{\alpha^S} =0.9$ ,  $\rho_{\alpha^F} =0$,  $\eta_{\alpha^S} =1$  and $\eta_{\alpha^F} =3$ for example):
\begin{equation}
\begin{split}
&\alpha^S_t=\rho_{\alpha^S} \alpha^S_{t-1} + \eta_{\alpha^S} N^{\alpha^S}_t \\
&\alpha^F_t=\rho_{\alpha^F} \alpha^F_{t-1} + \eta_{\alpha^F} N^{\alpha^F}_t \\
&\alpha_t=\alpha^S_t+\alpha^F_t
\end{split}
\end{equation}
where $N^{\alpha^S}_t$ and $N^{\alpha^F}_t$ are independent standard normal variables.\\

\noindent This model will enable us to take into account the interaction between fast and slow return predictors and the last weight in the optimal action.

\subsubsection{Risk modeling}

Let us now state the two risk terms that we will consider: $L_2$ and max weight

\paragraph{$L_2$ risk}

This risk term is equal to a risk aversion times the square of the position:
\begin{equation}
 \text{Risk}(w_t) = \frac{\lambda}{2} w_t^2.
\end{equation}

\noindent Note that this risk term corresponds to penalizing the variance of a portfolio holding $w_t$ in the stock.

\paragraph{Max weight as a risk}

As explained in \cite{chaouki2020deep}, we can impose a max weight constraint as a penalization of weights above a threshold $M$ (or below $-M$). Here we choose a $ReLU$ penalization:
\begin{equation}
\text{Risk}(w_t) = K(|w_t|-M)_+
\end{equation}
where $K$ is large so that the optimal weight is never above $M$.

\subsubsection{Cost modeling}

Let us now state the two cost terms that we will consider: $L_2$ and $L_1$.

\paragraph{$L_2$ cost} 

As in \cite{garleanu2013dynamic}, the cost term can be modeled as the square of the turnover. This enables to have closed form solutions but is not very realistic:
\begin{equation}
\text{Cost}(w_t, lw_t) = C\times|w_t-lw_t|^2.
\end{equation}

\paragraph{$L_1$ cost}

To take into account spread costs, we can model the cost as proportional to the turnover:
\begin{equation}
\text{Cost}(w_t, lw_t) = S\times|w_t-lw_t|.
\end{equation}
\noindent Note that this function is not strictly speaking differentiable. However, it is differentiable almost everywhere. We can thus apply SGD like methods to minimize the corresponding loss (in the same way that we can apply SGD to minimize the parameters of neural networks with $ReLU$ activation functions).

\section{Numerical results}

\label{sec:num_res}

In this section, we fit and describe the optimal action function in five trading environments:

\begin{itemize}
\item{An environment with a mono-scale alpha, a $L_1$ cost and a $L_2$ risk to show that we recover the result of \cite{chaouki2020deep} and study the behavior of our algorithm.}
\item{An environment with a mono-scale alpha, a $L_1$ cost and a max weight risk to show that we recover the result of \cite{chaouki2020deep} in a more numerically difficult case where the optimal action is not a continuous function of alpha.}
\item{An environment with a mono-scale alpha with a variable alpha scaling, a $L_1$ cost and a $L_2$ risk. With this environment we want to prove that our algorithm can find the optimal strategy for a continuum of environment parameters with only one fit and  study the impact of the alpha variability on the optimal strategy.}
\item{An environment with a multi-scale alpha, a $L_1$ cost and a $L_2$ risk to study the impact of the fast alpha on the optimal strategy.}
\item{An environment with a multi-scale alpha, a $L_1$ cost with a variable spread and a $L_2$ risk to study the impact of the spread on the effect of the fast alpha on the optimal strategy.}
\end{itemize}

\subsection{Mono-scale alpha}
\label{subsec:mono}

To study the convergence of the algorithm towards the optimal action we begin by a simple environment where:
\begin{itemize}[noitemsep]
\item The alpha is mono-scale with persistence parameter $\rho_\alpha=0.9$ and scaling $\eta_\alpha=1$.
\item There is a $L_1$ cost term with spread $S=4$.
\item There is a $L_2$ risk term with risk aversion $\lambda=1$.
\end{itemize}


\noindent There are different behaviors depending on the model parameters: If the spread is too high compared to the amplitude of the expected return, then the optimal strategy is to (almost) never trade. If the spread is too low, the optimal strategy is to (almost) always trade. With the parameters chosen above, we are between the two regimes: we trade about once every 5 periods (see Figure  \ref{fig:weight_alpha_trajectory_mono} below).

\subsubsection{Properties of the optimal action}

\begin{figure}[H]
\centering
\includegraphics[width=\textwidth]{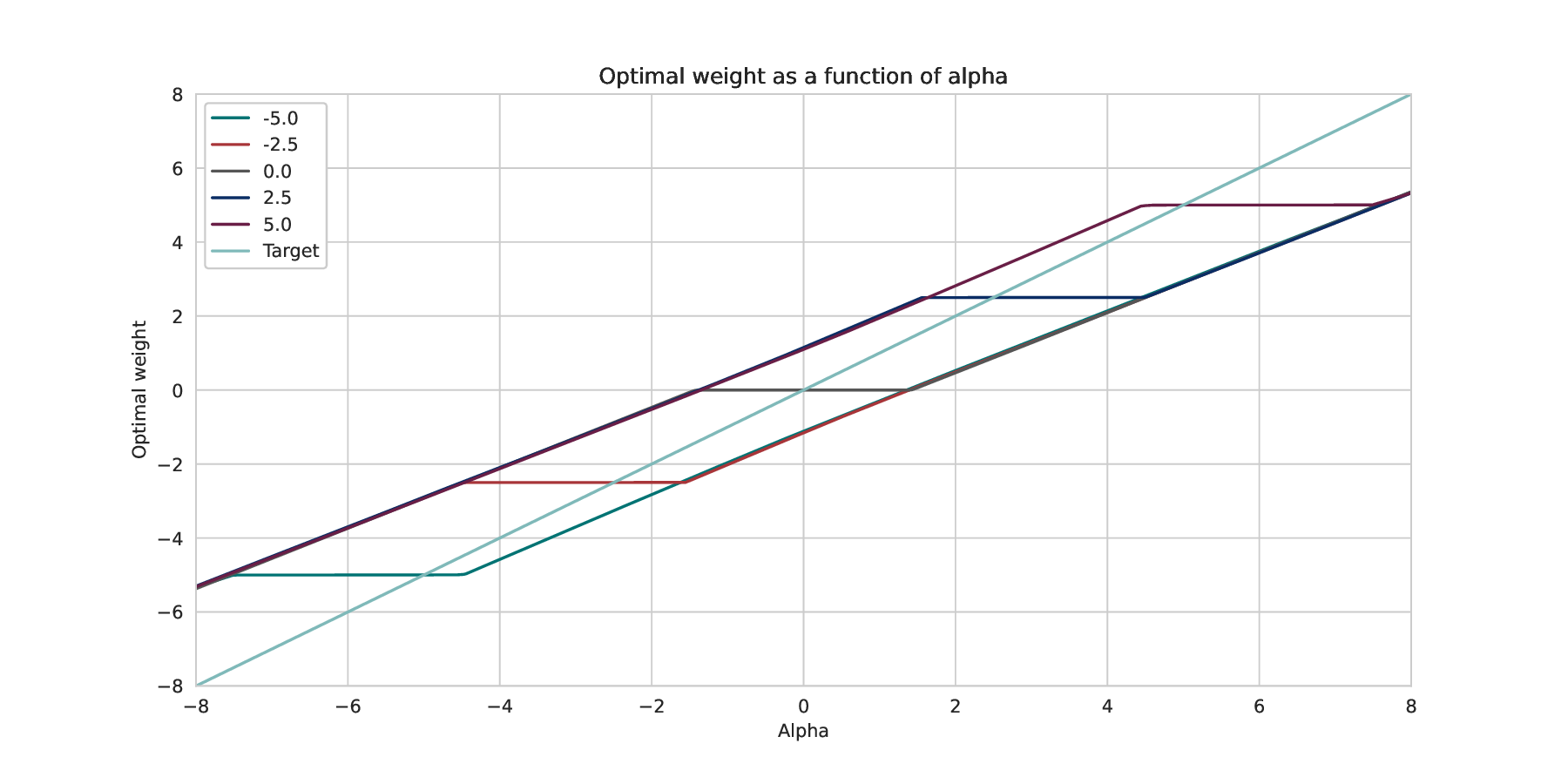}
\caption{Optimal weight as a function of the current alpha for different last weights and the target weight which in this case is the alpha.}
\label{fig:weight_of_alpha_mono}
\end{figure}

The optimal weight is a function of the last weight and the alpha:
$$a_t=A(\alpha_t, lw_t).$$
 We represent it by plotting for a few last weight values the optimal weight as a function of alpha in Figure \ref{fig:weight_of_alpha_mono}.

\noindent Note that in this model, if there is no spread costs, the optimal weight is the one which maximizes the current reward $r_t$ which is $w^{tgt}_t=\alpha_t / \lambda$. In the current case where $\lambda=1$, $\alpha$ can thus be seen as a target weight. We find as in \cite{de2012optimal} that the optimal strategy has the same functional form as in the single period case, see \cite{dybvig2020mean}, and consists in not trading if the alpha is within a threshold around the last weight and to get to this threshold if the alpha is outside of it.

\begin{figure}[H]
\centering
\includegraphics[width=\textwidth]{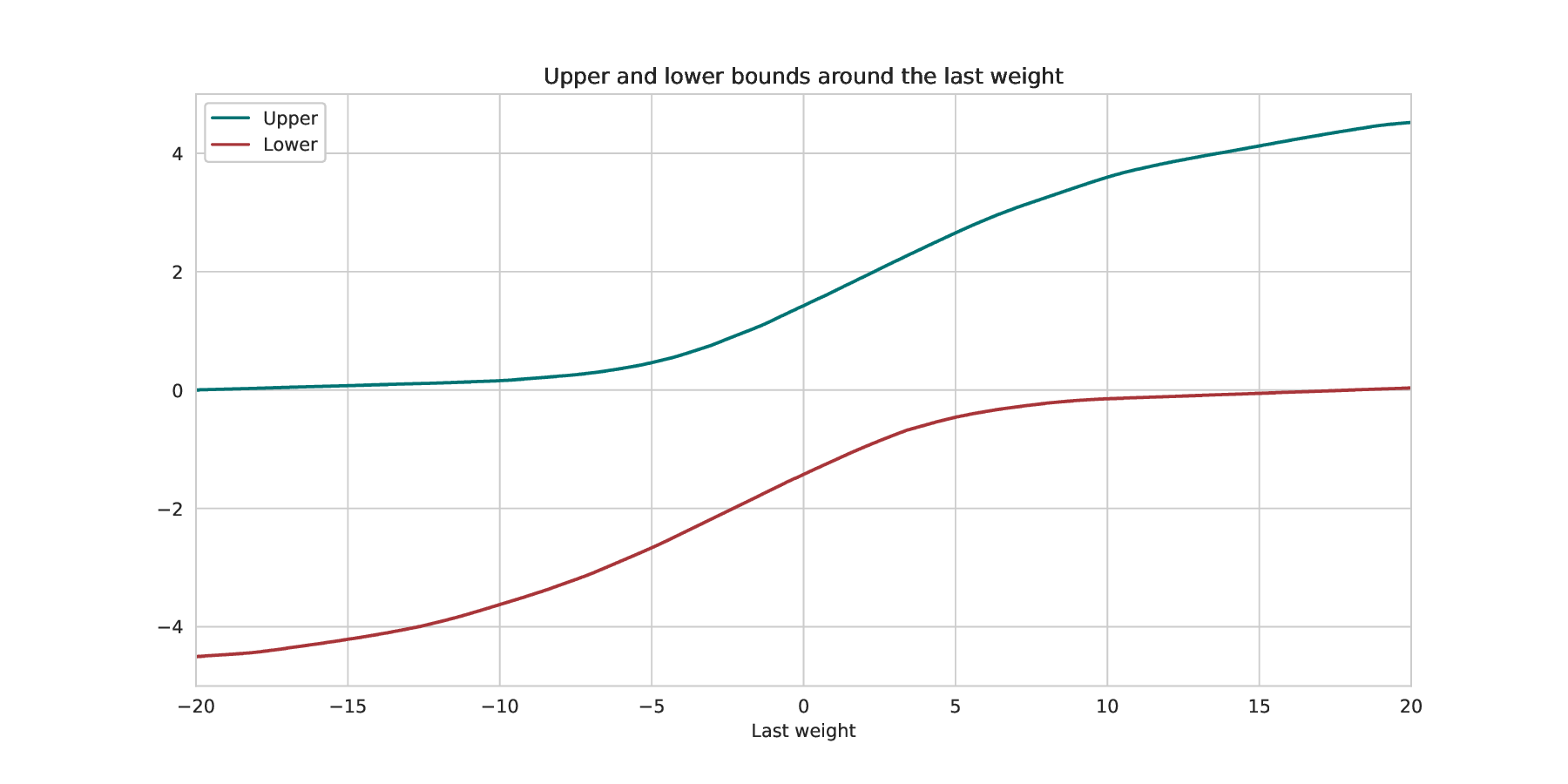}
\caption{Upper and lower trading bounds as a function of the last weight.}
\label{fig:trading_bounds}
\end{figure}

\noindent These thresholds are analyzed theoretically in \cite{de2012optimal}. We plot them in Figure \ref{fig:trading_bounds}. The upper (resp. lower) bound is defined as the smallest (resp. largest) alpha such that the optimal weight is strictly above (resp. below) the last weight to which we subtract the last weight. We find that except for a zero last weight, this threshold is asymmetric: for positive (resp. negative) last weight, the upper bound in larger (resp. smaller) than minus the lower bound. For high (resp. low) last weights, the lower (resp.upper) bound tends to zero. This means that for a high (resp. low) last weight, the optimal strategy is to sell (resp. buy) as soon as the alpha/target weight is below (resp. above) the last weight.\\

\begin{figure}[H]
\centering
\includegraphics[width=\textwidth]{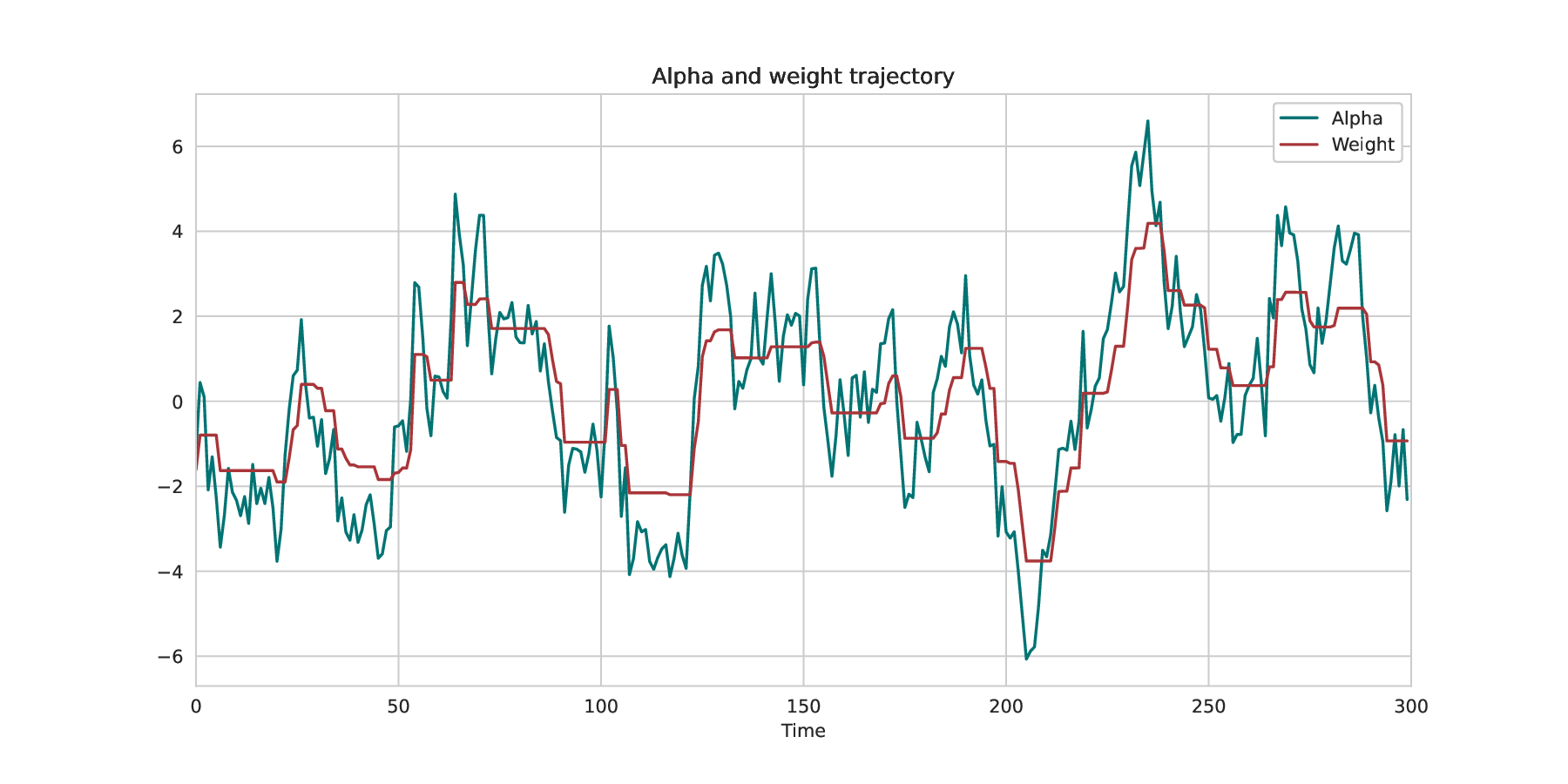}
\caption{Sample of the trajectory of the alpha signal and optimal weight.}
\label{fig:weight_alpha_trajectory_mono}
\end{figure}

\noindent We plot in Figure \ref{fig:weight_alpha_trajectory_mono} a sample trajectory of the alpha and of the associated optimal weight. We see that when the alpha gets too far from the weight, trading happens.

\subsubsection{Convergence speed}

\begin{figure}[H]
\centering
\includegraphics[width=\textwidth]{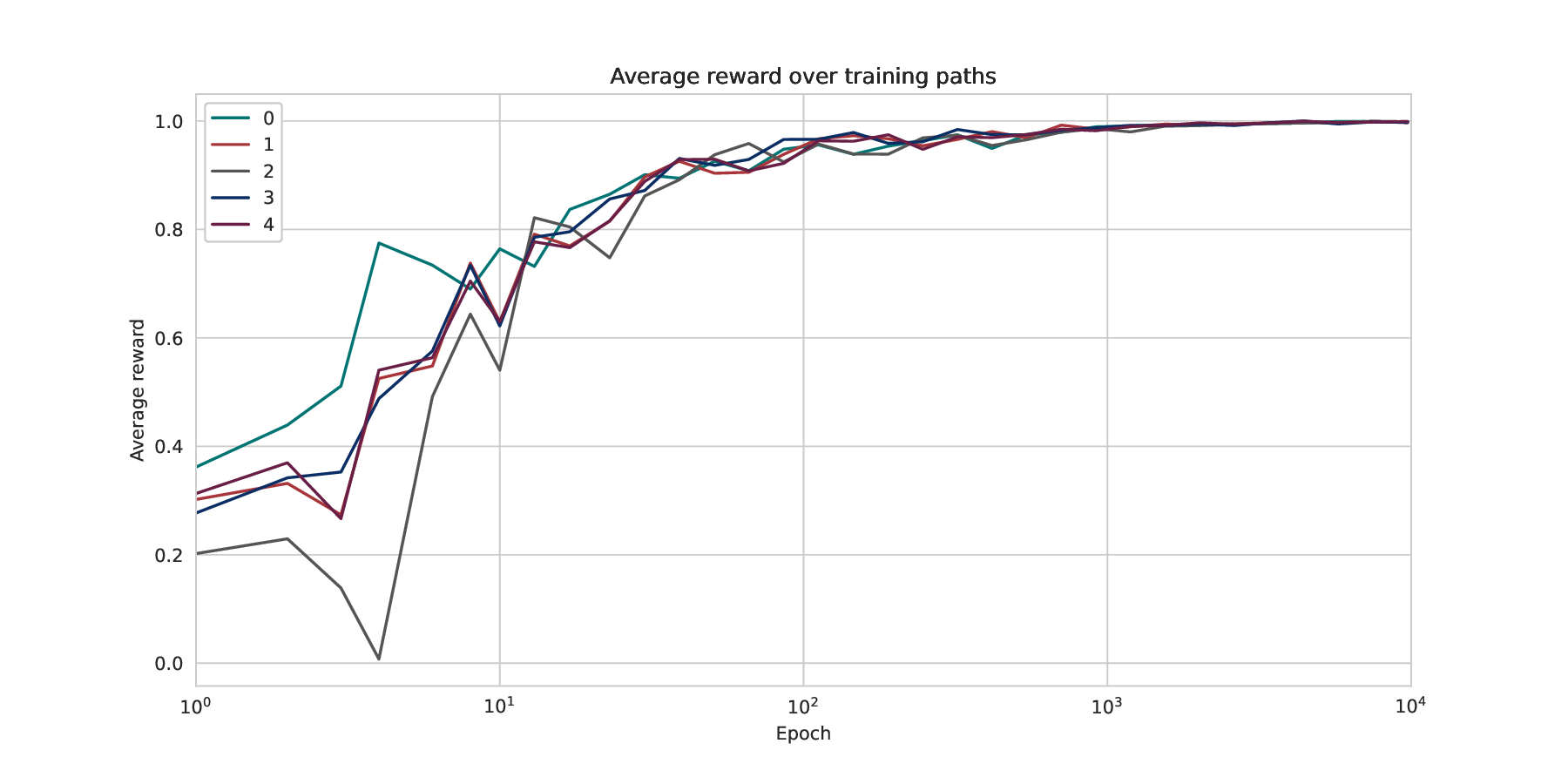}
\caption{Evolution of the validated (on 1 million samples) reward over 5 training paths. The x axis corresponds to the number of mini-batches (each having 1024 samples).}
\label{fig:TrainingPaths}
\end{figure}

To showcase the stability of the convergence of the algorithm we plot in Figure \ref{fig:TrainingPaths} the evolution of the average  reward (normalized to one for the optimal strategy) through learning paths over the first epoch of the optimization.

\noindent We find that the randomness of the starting point and of the sampling does not affect the obtained strategy as it is sometimes the case when fitting deep neural networks. We thus do not need to train multiple agents and take the best.\\

\noindent Note also that  after 1000 1024 minibatches, corresponding to one epoch over 1 million samples, the validation cumulated reward is above 99\% of the optimal cumulated reward for 4 of the 5 training paths. Therefore, taking 50 epochs on 10 million samples is very conservative.

\subsubsection{Impact of the horizon $T$}

In this paragraph, we study the impact of the horizon $T$ on the obtained strategy. Intuitively, this horizon should be longer than the `memory' of the model. For example, here the autocorrelation of the alpha process behaves as $(\rho_\alpha)^t$ and thus, the half-life of the model is $-\log(2)/\log(\rho_\alpha)\sim 6.5$. Hence, one should take a horizon of at least 10.
Plotting in Figure \ref{fig:RewardOfHorizon} the long term reward as a function of the horizon, this is roughly what we find: The reward (and the associated strategy) converges around $T=5$. Recall that to be conservative and allow for longer memory models, we take $T=50$.

\begin{figure}[H]
\centering
\includegraphics[width=\textwidth]{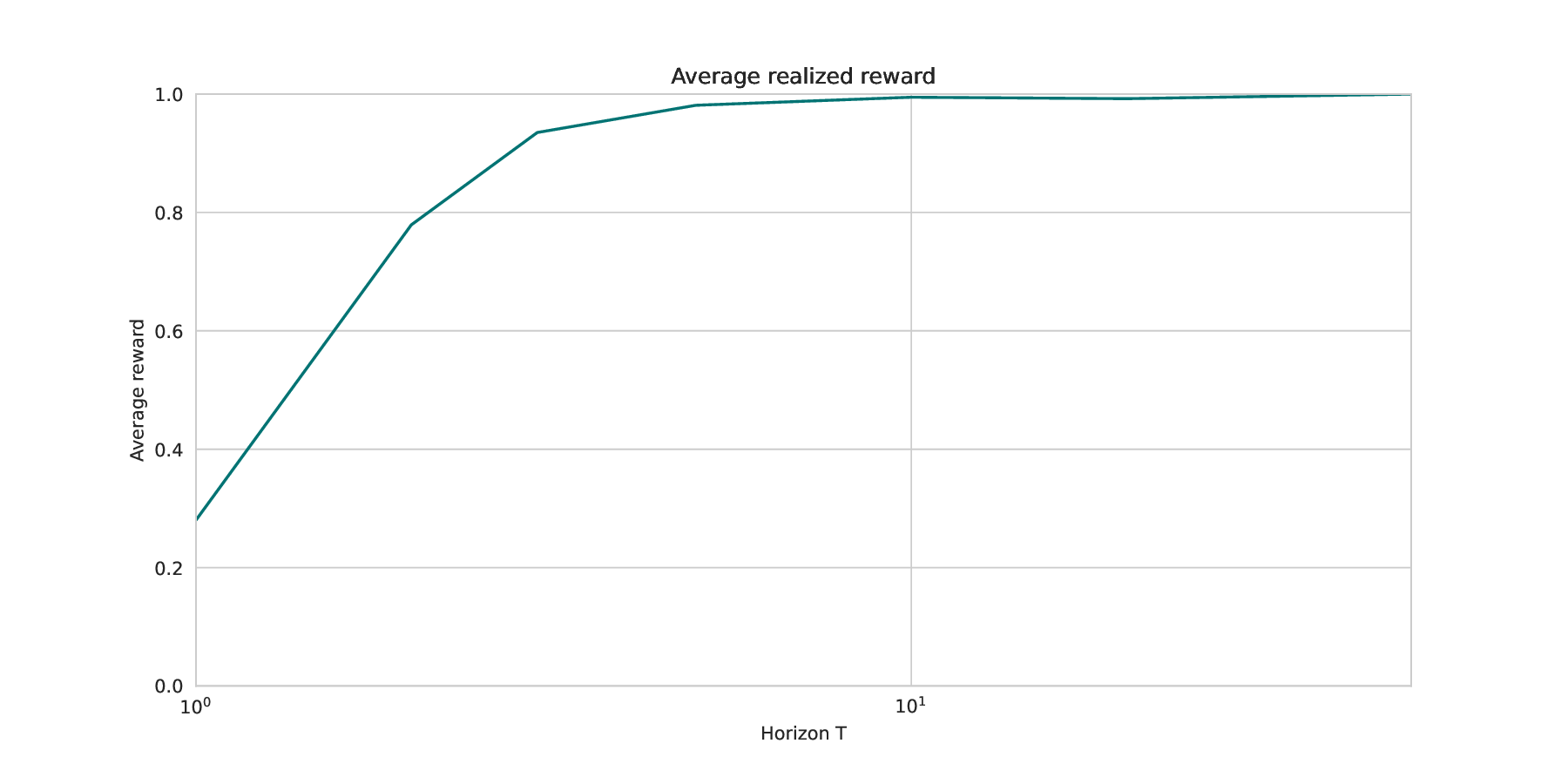}
\caption{Average long term reward as a function of the horizon $T$.}
\label{fig:RewardOfHorizon}
\end{figure}

\subsection{Max weight risk}

An environment which is slightly harder to solve numerically, because of the discontinuity of the optimal action as a function of alpha, is the following:

\begin{itemize}[noitemsep]
\item As before, the alpha is mono-scale with persistence parameter $\rho_\alpha=0.9$ and scaling $\eta_\alpha=1$.
\item There is a $L_1$ cost term with spread $S=4$.
\item There is a max weight risk term with max weight $M=3$ and penalization $K=10$ (but no $L_2$ risk aversion).
\end{itemize}

\begin{figure}[H]
\centering
\includegraphics[width=\textwidth]{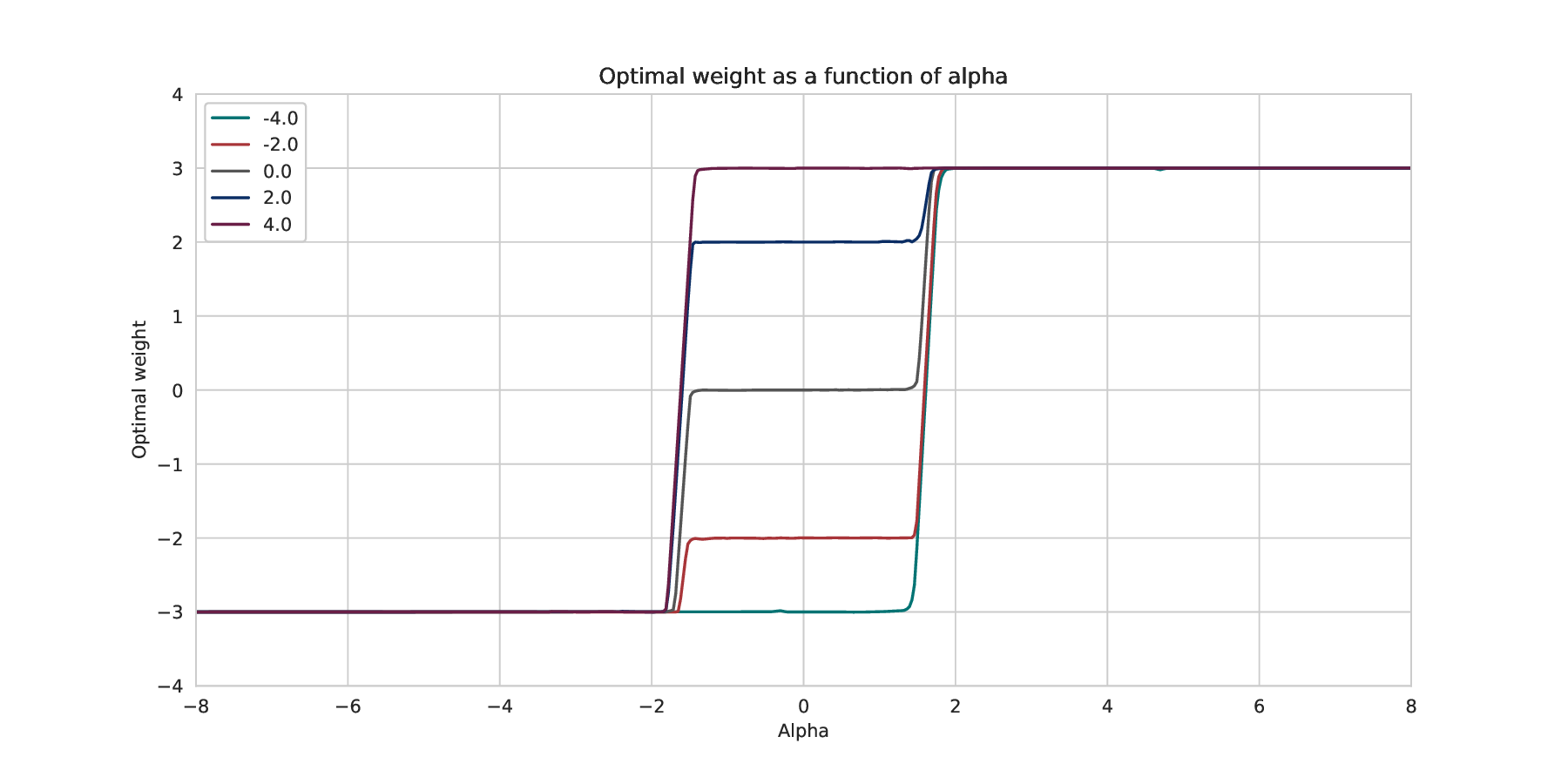}
\caption{Optimal weight as a function of the current alpha for different last weights.}
\label{fig:max_weight_no_risk_aversion}
\end{figure}

\noindent We plot in Figure \ref{fig:max_weight_no_risk_aversion} the optimal weight as a function of alpha for a few last weights as above. We find as in \cite{chaouki2020deep} that the optimal solution is to get to the max weight (resp. minus the max weight) if the alpha is greater (resp. lower) than a limit and not to trade between. The optimal action as a function of alpha is thus discontinuous. Although the fit is less good than before (because a neural network needs large parameters to get large derivatives) it still captures the global shape of the optimal strategy. Note that if we increase the number of epochs, we get closer and closer to the perfect discontinuous real solution.

\subsection{Alpha scaling impact}

To show the ability of our model to treat environment parameters as static state dimensions, we train a meta-model which finds the optimal action as a function of the last weight, the alpha signal and the alpha scaling:$$a_t=A(\alpha_t, lw_t,\eta_\alpha).$$ To do this, we take the same model as in the first environment except that we sample the alpha scaling uniformly over $[0, 4]$ and treat is as a static (that is non time varying) state dimension.\\

\noindent In Figure \ref{fig:RewardOfHorizon} we plot the optimal action function as a function of the alpha for a last weight equals to zero and for different values of alpha scaling.

\begin{figure}[H]
\centering
\includegraphics[width=\textwidth]{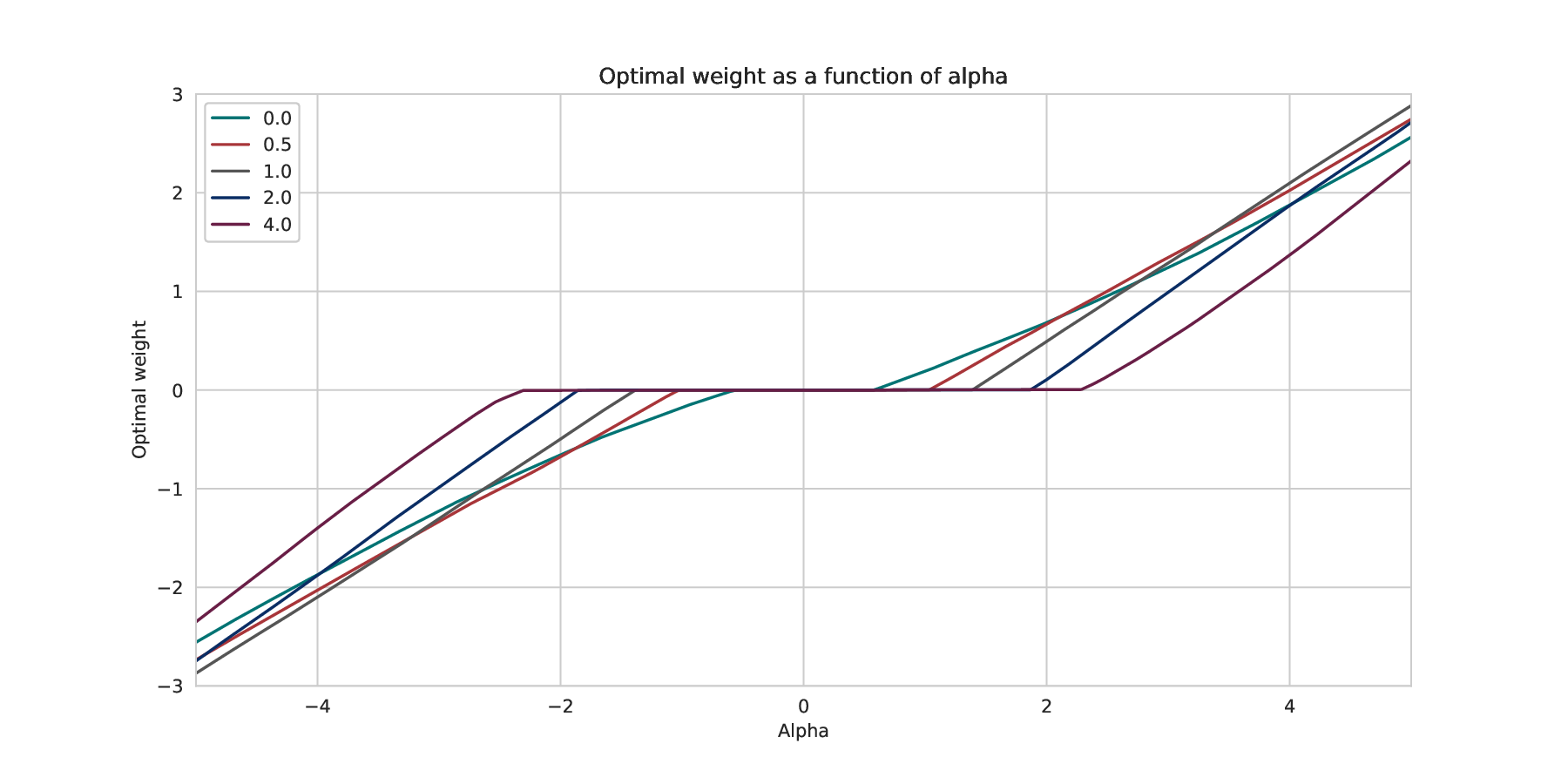}
\caption{Optimal weight as a function of the current alpha for different alpha scaling and last weight equals zero.}
\label{fig:RewardOfHorizon}
\end{figure}

\noindent Interestingly, we see that the optimal strategy strongly depends on the alpha variability. It makes sense that the non trading region increases with the alpha scaling. Indeed, if the alpha variability is large, it is more likely that the alpha will change sign shortly and one will need to revert his trades. The heuristic solution of \cite{boyd2017multi} which assumes that the future alpha trajectory is deterministic equal to its expectation corresponds to a alpha variability $\eta_\alpha=0$ (an $AR(1)$ process with no noise is equal to its expectation). In this framework, the obtained strategy is thus the curve $0$ in Figure \ref{fig:RewardOfHorizon}. Since in practise, the future alpha trajectory is uncertain ($\eta_\alpha>0$), we see that this heuristic can be significantly suboptimal. In particular, the no trading regions are smaller than they should be.

\subsection{Two-scale alpha}
 \label{subsec:2scales}

In this paragraph, we introduce the first multi-scale alpha environment. To our knowledge, outside of the case of $L_2$ transaction costs, the study of optimal strategies for a multi-scale alpha signals has never been done.

\begin{itemize}[noitemsep]
\item The alpha is two-scale with persistence parameters $\rho_{\alpha^S}=0.9$  and $\rho_{\alpha^F}=0$ and volatilities $\eta_{\alpha^S}=1$ and $\eta_{\alpha^F}=3$.
\item There is a $L_1$ cost term with spread $S=4$.
\item There is a $L_2$ risk term with risk aversion $\lambda=1$.
\end{itemize}

\noindent The parameters are chosen so that we are in a regime where the optimal weight has the same form as the alpha trajectory without trading at every period. The fast alpha magnitude is chosen such that, as it is the case in practice for large portfolios, it is rarely worth doing a round trip to exploit the fast alpha. There is however room to use the fast alpha to time the execution of the slow alpha.\\

\begin{figure}[H]
\centering
\includegraphics[width=\textwidth]{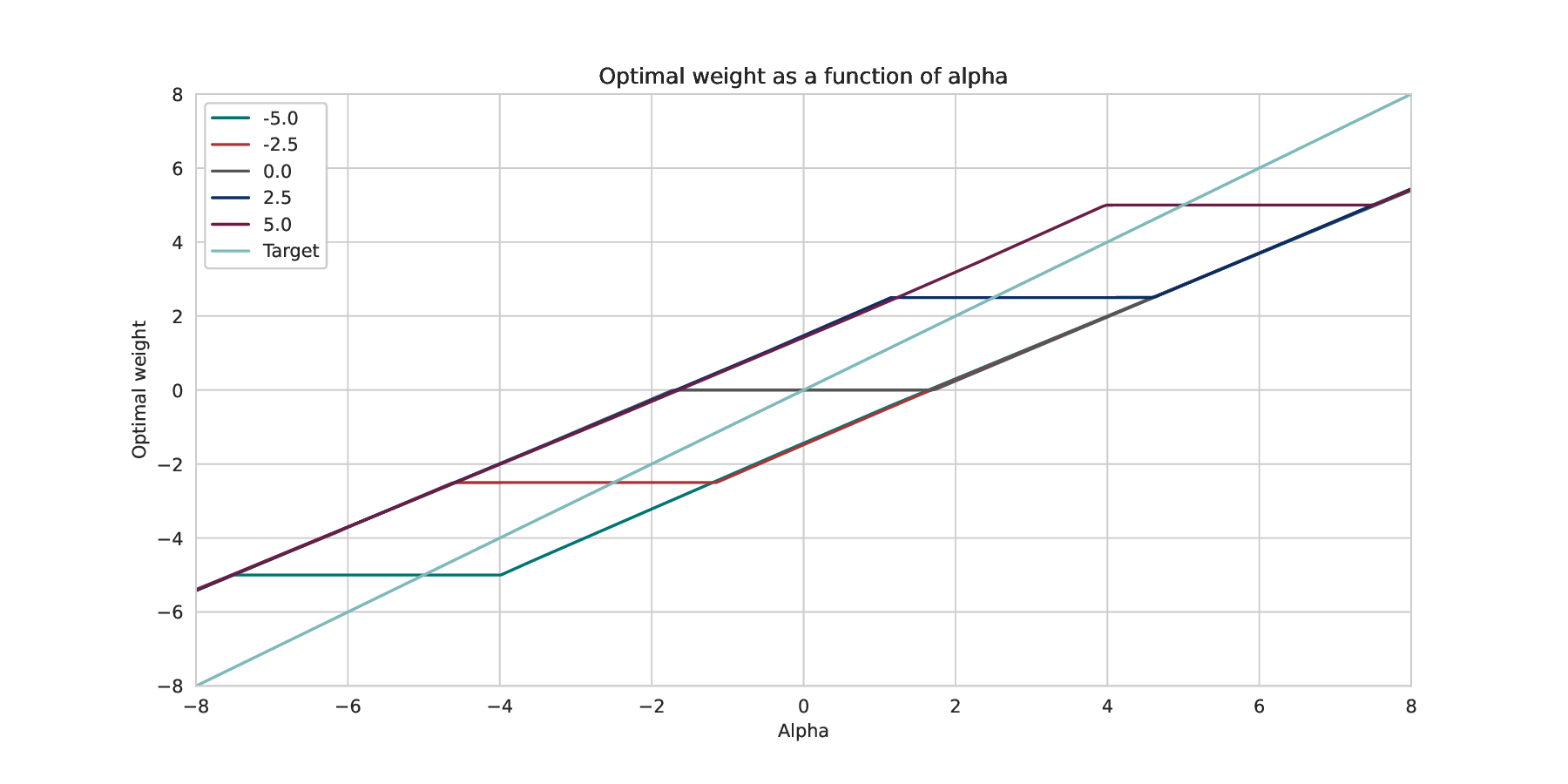}
\caption{Optimal weight as a function of the slow alpha for different starting weight and fast alpha equals zero.}
\label{fig:weight_of_alpha_2s}
\end{figure}

\noindent The optimal action is a function of the last weight, the slow alpha and the fast alpha:$$a_t=A(\alpha^F_t, \alpha^S_t, lw_t).$$To show the properties of the optimal strategy, we begin by fixing the fast alpha to zero and as before to plot in Figure \ref{fig:weight_of_alpha_2s} the optimal weight as a function of the slow alpha for different starting weight.

\begin{figure}[H]
\centering
\includegraphics[width=\textwidth]{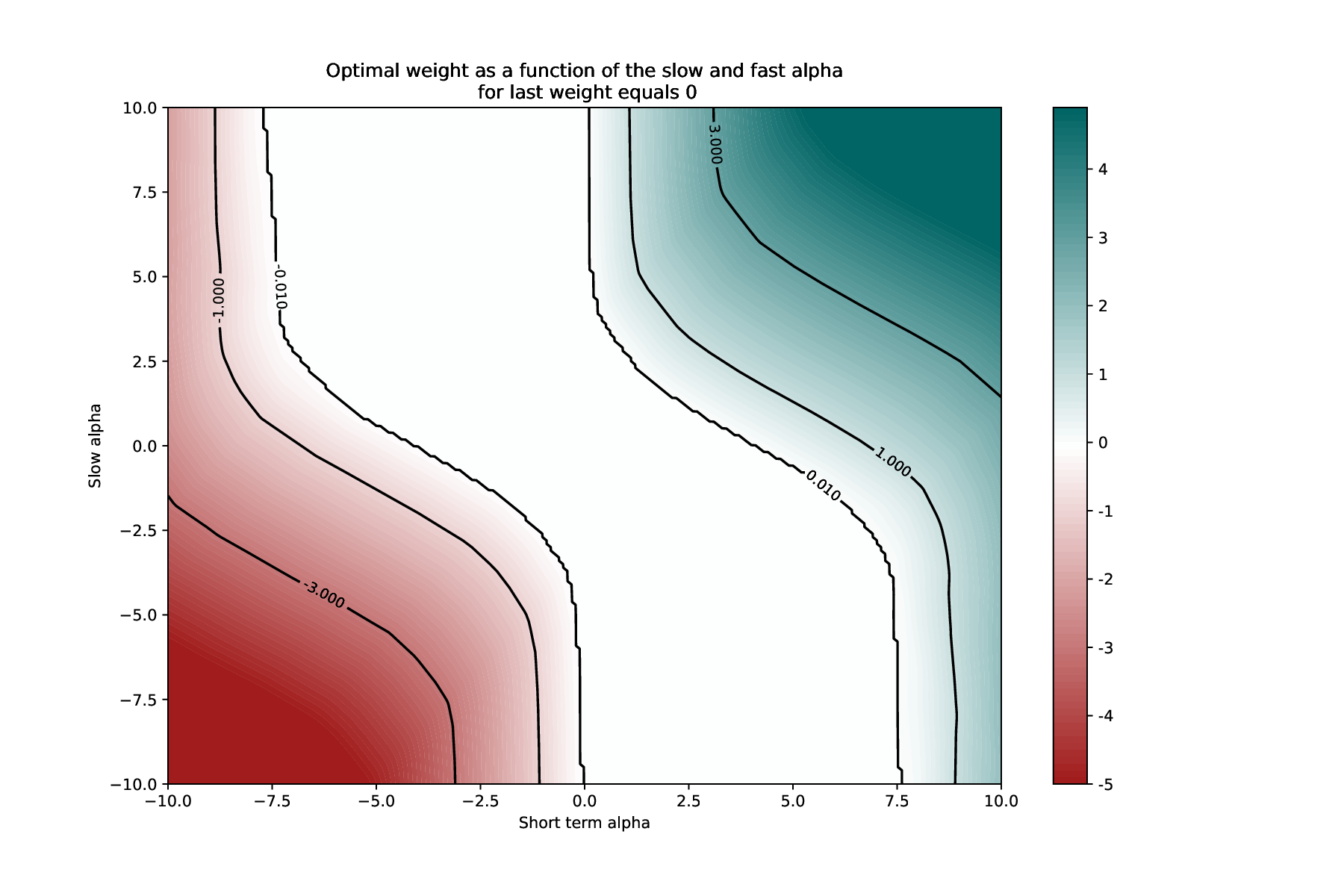}
\caption{Heat-map of the optimal weight as a function of the slow and fast alphas for the last weight equals zero.}
\label{fig:2sHeatmapLW0}
\end{figure}

\noindent We find the same behavior as in the first environment without the fast alpha. Let us now study the impact of the fast alpha on the optimal strategy. To do this, in Figure \ref{fig:2sHeatmapLW0}, we fix the last weight equals to zero and plot as a heat-map the optimal weight as a function of the slow alpha and the short term (or total)  alpha defined as the sum of the slow and fast alphas.\\

\noindent We see that this heat-map is split into three areas: 
\begin{itemize}[noitemsep]
\item On the bottom left, the optimal action is to sell (more and more as you get lower and lefter).
\item On the top  right, the optimal action is to buy.
\item In the middle, there is a no trading region. 
\end{itemize}

\noindent If the short term alpha is higher (resp. lower) than 2 spreads (resp. minus 2 spreads) we always buy (even if we expect to revert the trade next month). If the the short term alpha is higher (resp. lower) than zero we never sell (resp. buy) but prefer to wait next month.

\begin{figure}[H]
\centering
\includegraphics[width=\textwidth]{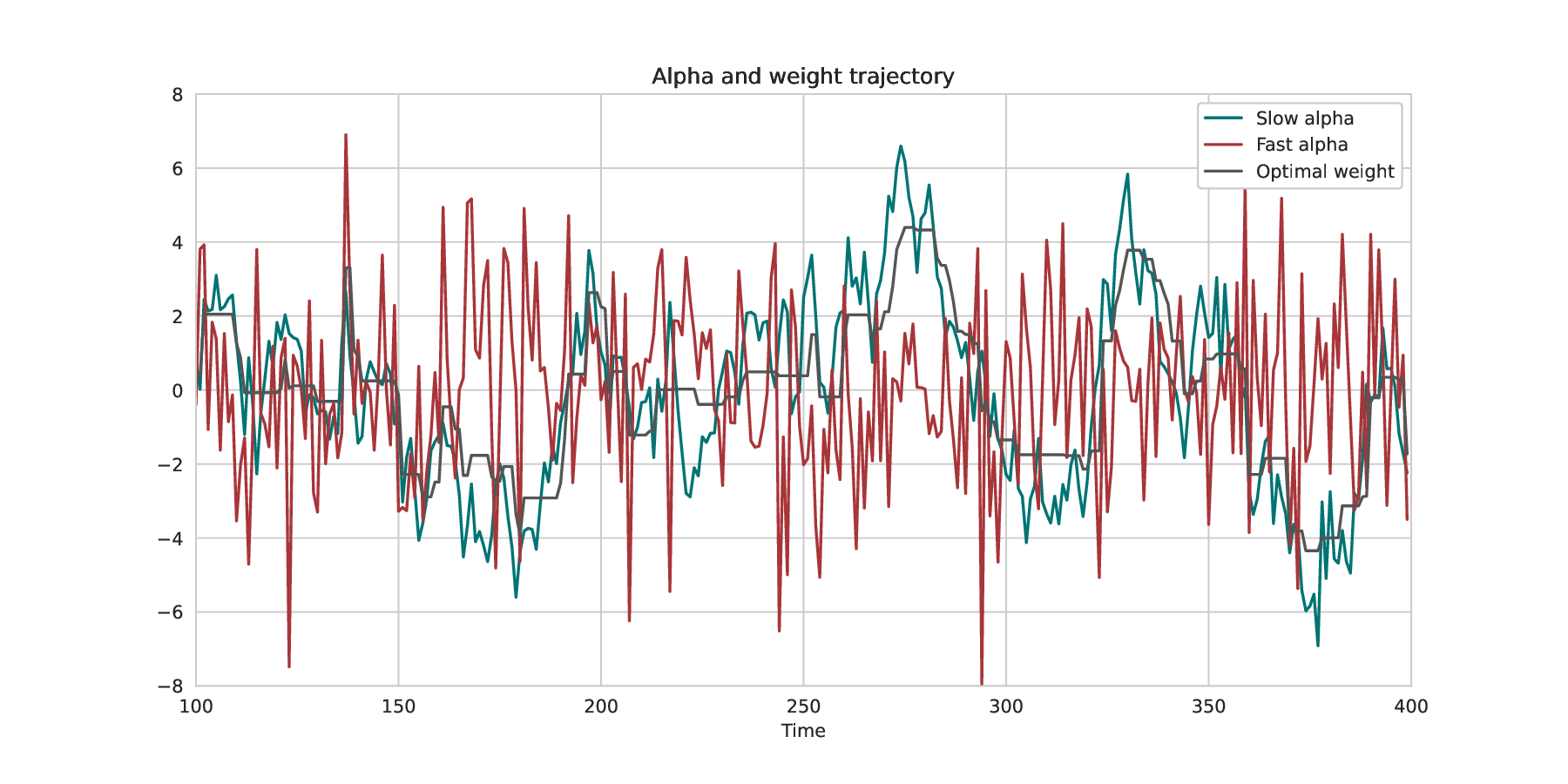}
\caption{Sample trajectories of the slow and fast alphas and of the optimal weight.}
\label{fig:weight_alpha_trajectory_2s}
\end{figure}

\noindent Looking at sample trajectories of the slow and fast alphas and of the optimal weight in Figure \ref{fig:weight_alpha_trajectory_2s}, we see that the main long term driver of the weight is the slow alpha and the fast alpha does not seem to matter much. However, zooming in, we see that the fast alpha is used to time the trades corresponding to the slow alpha.
Indeed, looking at the two blue ovals on Figure \ref{fig:zoom_weight_alpha_trajectory_2s} below, we see that at some points the slow alpha goes up but the optimal weight does not follow before the fast alpha confirms that it is worth trading now and not to wait. This is a quite intuitive result: If you know that a stock is going to perform well next year but not next month, you wait before buying it. However, if you already have the stock then you do not necessarily sell it.

\begin{figure}[H]
\centering
\includegraphics[width=\textwidth]{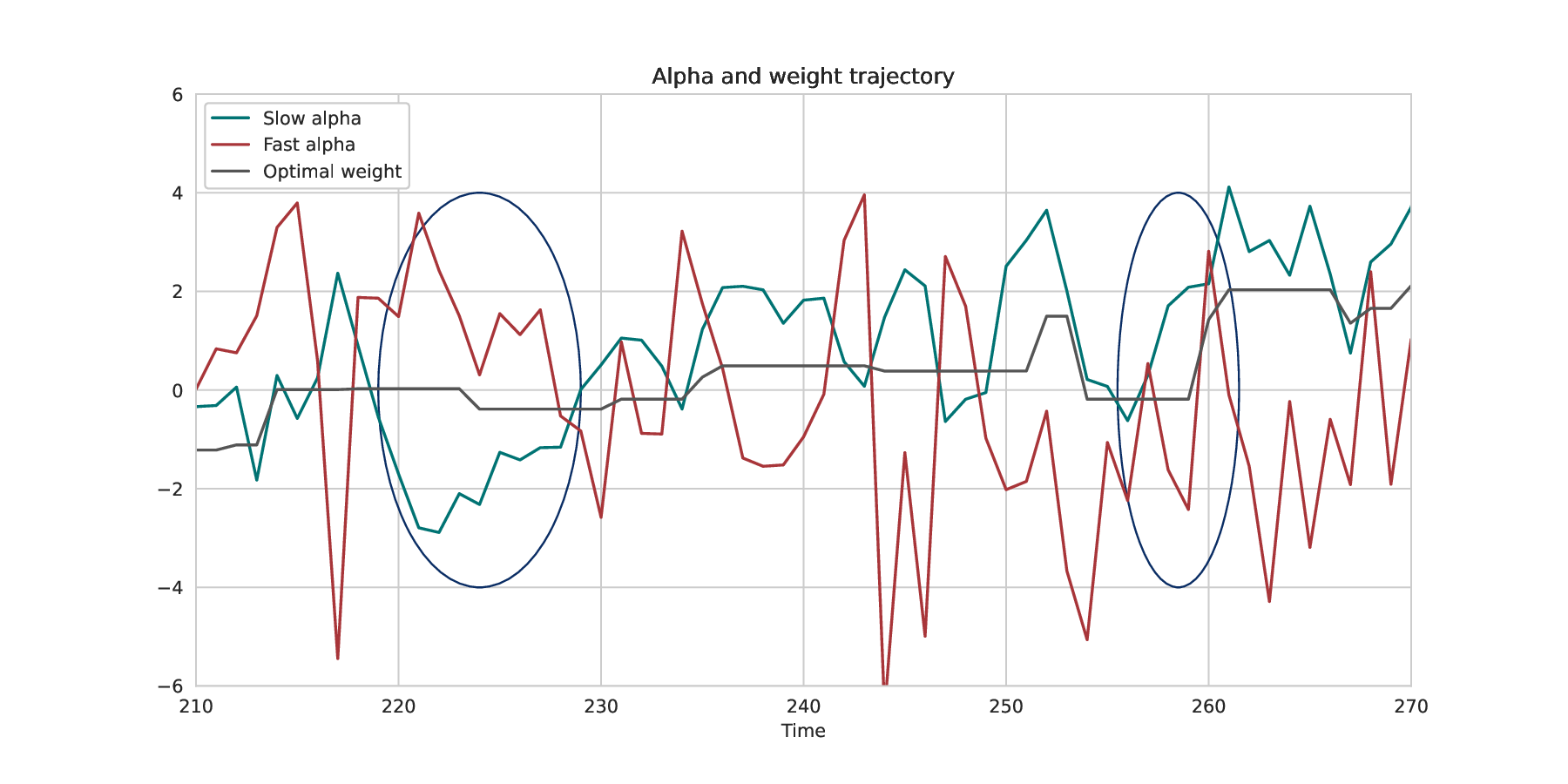}
\caption{Zoom on sample trajectories of the slow and fast alphas and of the optimal weight.}
\label{fig:zoom_weight_alpha_trajectory_2s}
\end{figure}

\noindent We confirm this by computing the empirical correlations between alphas, optimal weights and trades (defined as $w_t-lw_t$). We find that although the slow alpha is strongly correlated to the optimal weight (88\%), trades are more correlated to the fast alpha (50\%) than to the slow alpha (26\%).

\begin{table}[H]
\centering

\begin{tabular}{lrrrr}
\toprule
{} &  Slow alpha &  Fast alpha &  Optimal weight &  Trade \\
\midrule
Slow alpha     &        1.00 &         &             &  \\
Fast alpha     &        0.00 &        1.00 &         &     \\
Optimal weight &        \textbf{0.88} &        0.17 &            1.00 &   \\
Trade          &        0.26 &        \textbf{0.50} &            0.17 &   1.00 \\
\bottomrule
\end{tabular}
\caption{Correlation matrix between weights, trades and slow and fast alphas. We see that although the weights are driven by the slow alpha, the trades are driven by the fast alpha.}
\end{table}

\subsection{Spread impact}

To study the impact of the spread value on the effect of the fast signal in the optimal strategy, we consider the same environment as in the previous paragraph except that we treat the spread as a static state variable. We sample the spread uniformly over [0, 6].\\

\begin{figure}[H]
\centering
\includegraphics[width=\textwidth]{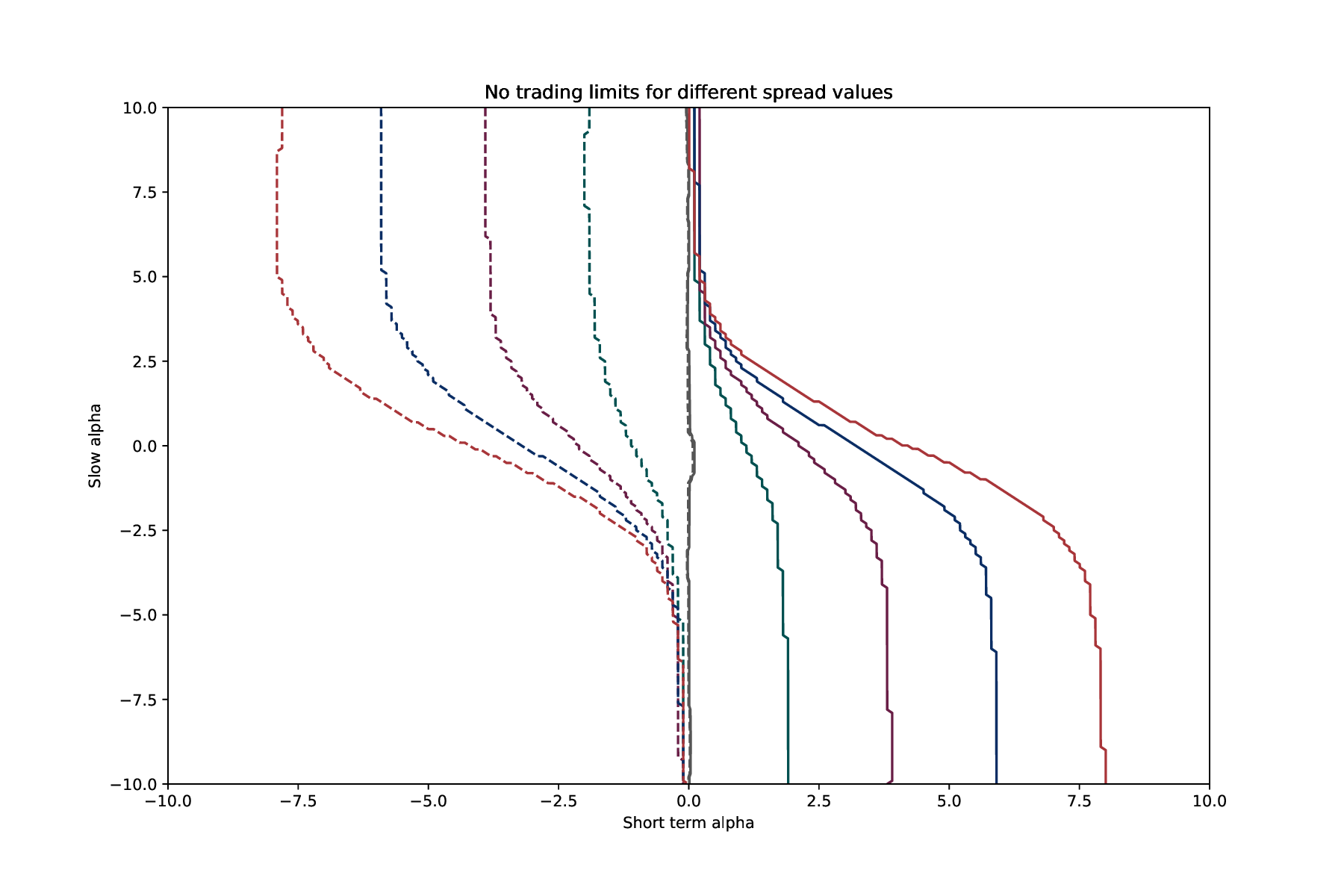}
\caption{No trading regions as a function of the slow and short term alpha for different spread values: 0 (gray), 1 (green), 2 (purple), 3 (blue) and 4 (red). On the left of dashed lines, we sell, on the right of full lines we buy and between the two we do not trade.}
\label{fig:weight_of_fastalpha_2s}
\end{figure}

\noindent In Figure \ref{fig:weight_of_fastalpha_2s} we plot the no trading zones for 4 spread values.

\noindent As expected, the higher the spread the larger the no trading zone and when the spread is zero the no trading zone is empty. We also find that as the spread decreases, conditionally on the short term alpha, the slow alpha matters less and less and only the short term alpha is important. When the spread is zero, the contour lines are fully vertical and the slow alpha does not matter.

\section{Conclusion}

\label{sec:conc}

In this article, we introduced the differentiable reinforcement learning framework where the environment is not a black box but its transition and reward functions are differentiable functions of the state, the action and some randomness that can be sampled. We noted that many well studied reinforcement learning and optimal control problems fall into this framework.
Using this environment knowledge, we proposed a simple algorithm to devise optimal action functions by writing them as deep neural networks. The stability and speed of this algorithm allows us to work in complex continuous multidimensional state and action environments. We can also simultaneously find a solution for many environments with one training by considering model parameters as a static state variables.\\

\noindent We applied this framework to devising sequential optimal trading strategies. We started by reproducing some well known strategies when the alpha process has only one time scale. We then defined and tackled environments where the expected return is a combination of a slow and fast component and studied the interactions of these two components with the current weight.
We found that for reasonable parameters, the fast alpha is used to time the trades corresponding to the slow alpha. We also studied the impact of the model parameters such as the alpha scaling and the spread on the optimal strategy.\\

\noindent Going further, being able to have access to the optimal strategy for any model parameters allows us to extend this by combining these strategies over thousands of stocks.  One way to do this is to proceed iteratively by solving the problems for all stocks independently and then to include risk and constraint terms through penalization of the alphas of stocks.
Another research direction is to consider more complex models with even more alpha time scales and transient market impact to apply deep learning to get numerical solutions to the extensively studied optimal execution problems.

\printbibliography

\appendix

\section{Pseudo code of the algorithm}
\label{app:code}

\noindent In addition to the stability and precision improvement, another advantage of the deep differentiable reinforcement learning approach compared to more generic reinforcement learning ones is code simplicity. Using the Pytorch framework (or any other automatic differentiation frameworks such as Tensorflow)  the code is very simple and has little parameters which need to be tweaked.\\

\noindent The parameters of the algorithm are:
\begin{itemize}
\itemsep0em 
\item The horizon $T$ of the cumulated reward.
\item The number of samples $N$.
\item The number of epochs.
\item The decrease rate at each epoch of the learning rate $\gamma$.
\item The Adam parameters (batch size, learning rate, learning rates, betas) that we let to their default values.
\end{itemize}

\noindent It takes as input an environment object which must have 4 methods:
\begin{itemize}
\itemsep0em 

\item $e.transition(state, action, U)\mapsto state$.
\item $e.reward(state, action, V)\mapsto reward$.
\item $e.generate\_randomness(N, T) \mapsto (U, V)$.
\item $e.initialize()\mapsto state$.
\end{itemize}

\noindent We then need to implement the cumulated reward function as

\begin{algorithmic}
\Function {$CR_T$}{$s_0, (U_t)_{t\leq T}, (V_t)_{t\leq T}, A, e$}

\State $R\gets 0$
\State $a_0 \gets A(s_0)$
\For{$t \gets 0, T-1$}

\State $R+=e.reward(s_t, a_t, V_t)$
\State $s_{t+1} \gets e.transition(s_t, a_t, U_t)$
\State $a_{t+1} \gets A(s_{t+1})$
\EndFor

\Return {R}
\EndFunction
\end{algorithmic}

\noindent Once we have these ingredients, we simply need follow the procedure:

\begin{enumerate}
\itemsep0em 
\item Sample $N$ randomness samples $U$, $V$.
\item Initialize function $A$ as a dense feed forward neural network.
\item Define the average of the cumulated reward on these samples.
\item Minimize this average with respect to the parameters of $A$ using Adam.

\end{enumerate}

\section{Noised return reward}
\label{app:noise}

\noindent In this paragraph, we consider the environment of Section \ref{subsec:mono} except that as explained in Section \ref{subsec:reward} we take the noised return $R_{t+1}=\alpha_t + \sigma N^R_{t+1}$ instead of the alpha in the reward:
\begin{equation}
r_t=w_t R_{t+1} -  \text{Risk}(w_t) - \text{Cost}(w_t, lw_t).
\end{equation}
Note that this is framework corresponds to the case where the investor does not know a priori that the expected return is equal to $\alpha_t$ and needs to learn it from samples. We take the parameter $\sigma=50$ so that the scale of the signal to noise ratio is $\eta_\alpha/\sqrt{1 - \rho_\alpha^2}/\sigma$ is around 5\%.\\

\begin{figure}[H]
\centering
\includegraphics[width=\textwidth]{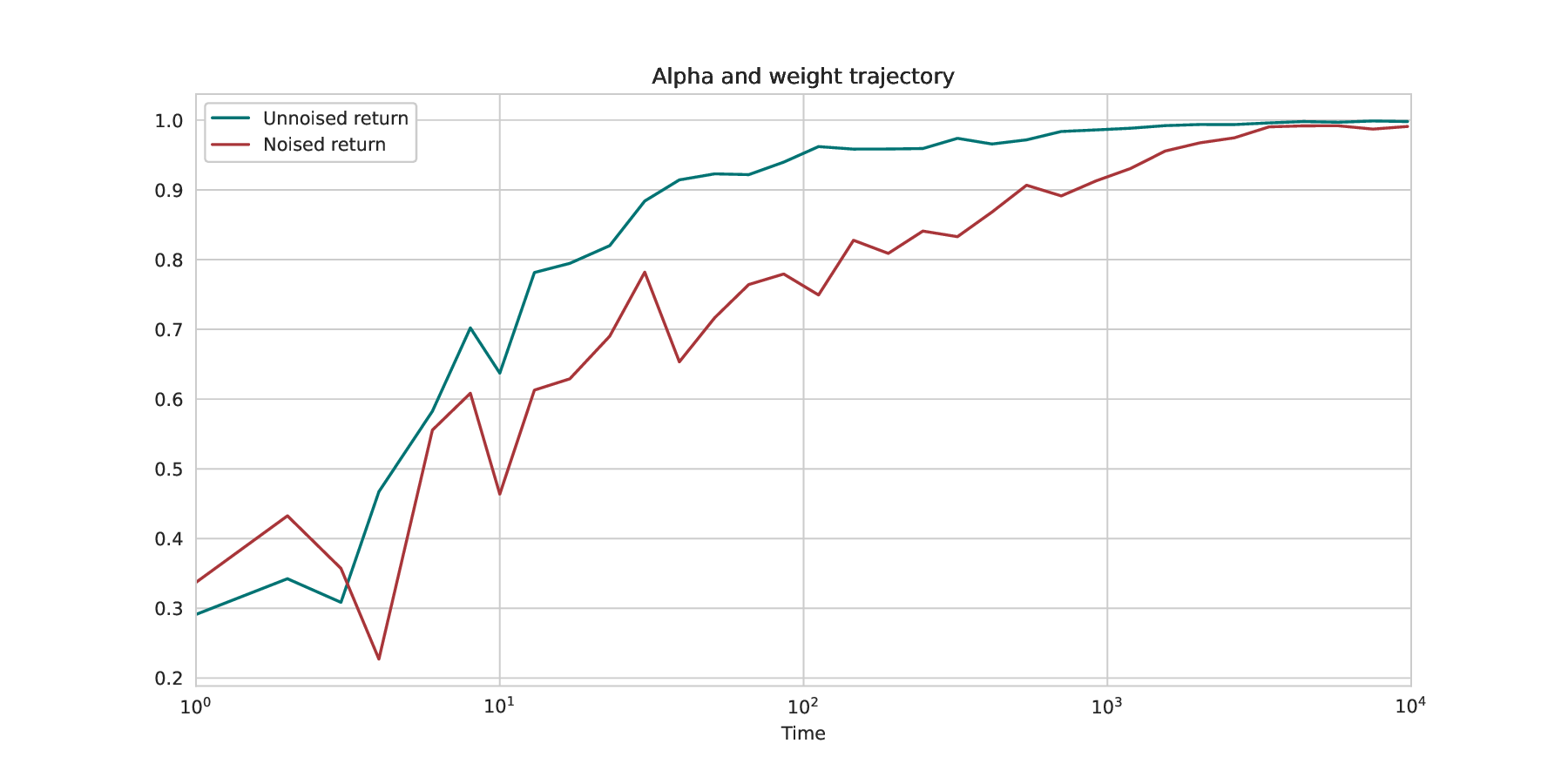}
\caption{Average validation (over 1 million samples) reward over 10 training paths of the algorithm applied to the environment of Section \ref{subsec:mono} versus the same one with noised return.}
\label{fig:noised_versus_unnoised}
\end{figure}

\noindent  In Figure \ref{fig:noised_versus_unnoised} we plot the average validation reward over 10 training paths of the algorithm applied to the environment of Section \ref{subsec:mono} versus the one defined above. We get that with noise, the algorithm takes longer to reach optimality: For example it takes 4000 steps instead of 1500 to reach 99\% of the optimal reward. However, after 50 epochs over 10 million samples, the obtained strategy is indistinguishable from the one without noise.

\section{Application of the DDPG algorithm}
\label{app:ddpg}

\noindent To showcase the stability and precision of the DDRL algorithm compared to more generic approaches, in this paragraph, we apply the DDPG algorithm of \cite{lillicrap2015continuous} (with the same parameters) to the two-scale alpha environment of Section \ref{subsec:2scales}. We find that through the learning path, the validation reward begins as expected by increasing and after about 15 000 mini batches starts to decrease going further from the optimal strategy.
\begin{figure}[H]
\centering
\includegraphics[width=\textwidth]{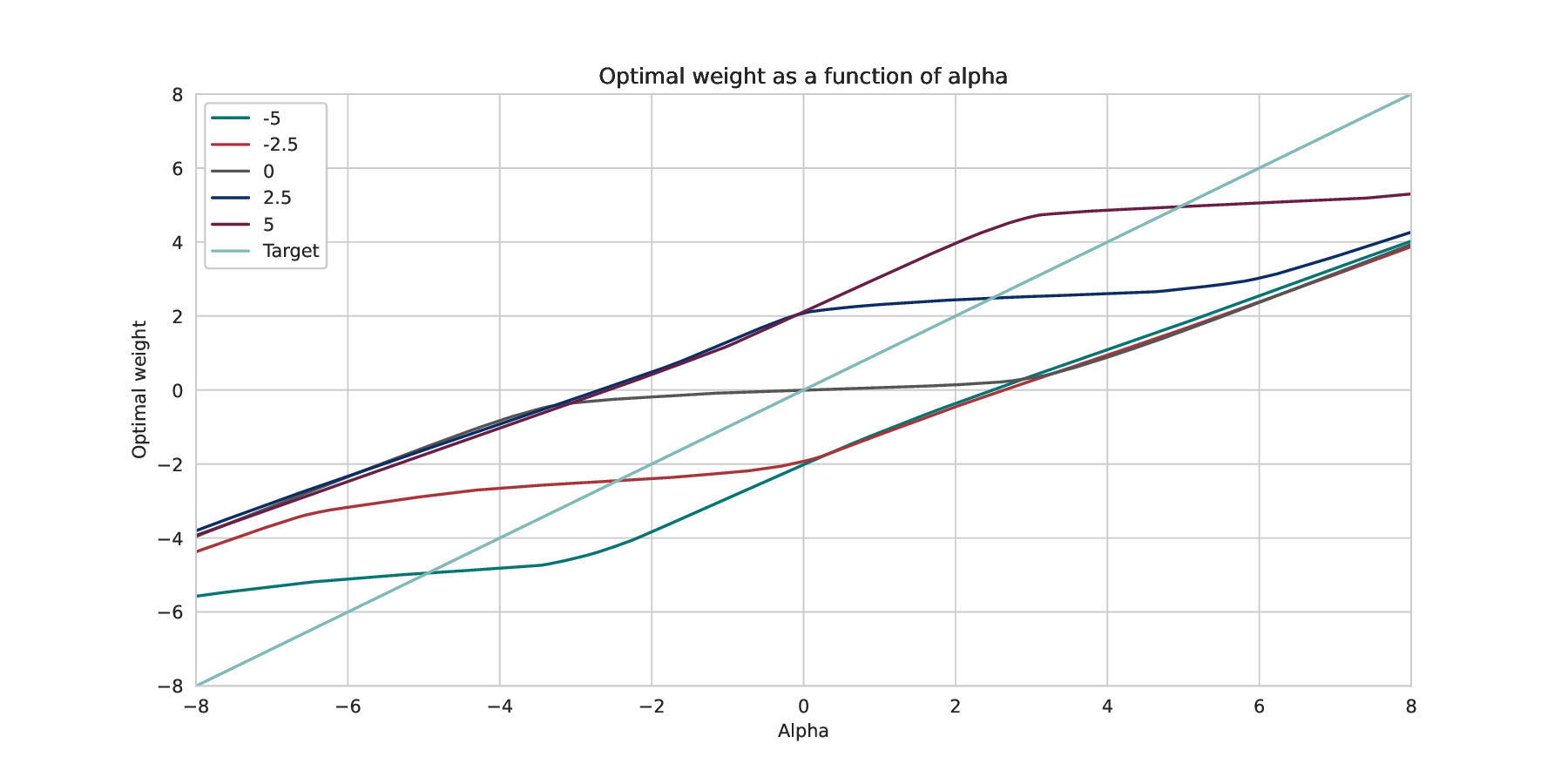}
\caption{Weight obtained with DDPG as a function of the slow alpha for different starting weight and fast alpha equals zero.}
\label{fig:weight_of_alpha_2s_ddpg}
\end{figure}

\noindent In Figure \ref{fig:weight_of_alpha_2s_ddpg}, we plot the action function  as a function of the slow alpha for different starting weight and fast alpha equals zero for the best agent of the learning path (in terms of validation reward over 1 million samples). This solution is unstable depending on the training path and lacks the symmetries of Section \ref{subsec:2scales}. It is therefore not surprising that its validation reward is only 87\% of that of Section \ref{subsec:2scales}. Tweaking the parameters and considering more complex variations of the algorithm such as twin delayed DDPG, see \cite{fujimoto2018addressing}, does not significantly improve the validated reward of the obtained solution.

\end{document}